\documentclass[manuscript,screen,nonacm,review=false]{jair}
\usepackage{amsmath,amsthm,booktabs,graphicx}
\RequirePackage[datamodel=acmdatamodel,style=acmauthoryear,backend=biber,
giveninits=true,uniquename=init]{biblatex}
\setcopyright{none}
\acmDOI{}
\acmVolume{}
\acmArticle{}
\acmYear{2026}
\acmMonth{10}
\JAIRAE{}
\JAIRTrack{}
\AtBeginDocument{\fancyfoot{}\fancyfoot[RO,LE]{\footnotesize Preprint, 18 October 2026}\fancypagestyle{firstpagestyle}{\fancyhf{}\fancyfoot[RO,LE]{\footnotesize Preprint, 18 October 2026}}}
\newtheorem{theorem}{Theorem}
\newtheorem{proposition}[theorem]{Proposition}

\newcommand{\bits}{\{0,1\}}

\newcommand{\F}{\mathbb F_2}
\begin{document}
\title[Pre-query retention with bit and word retrieval]{Pre-Query Retention with Bounded Bit and Word Retrieval}
\author{Pranav Singh}
\authornote{Both authors contributed equally to this work.}
\email{2023mcb1308@iitrpr.ac.in}
\affiliation{\department{Department of Mathematics}
\institution{Indian Institute of Technology Ropar}
\city{Rupnagar}\state{Punjab}\country{India}}
\author{Prashant Singh}
\authornotemark[1]
\email{2023mcb1309@iitrpr.ac.in}
\affiliation{\department{Department of Mathematics}
\institution{Indian Institute of Technology Ropar}
\city{Rupnagar}\state{Punjab}\country{India}}
\renewcommand{\shortauthors}{Pranav Singh and Prashant Singh}
\begin{abstract}
An agent can retain information before receiving a future question and
later retrieve evidence from an archive. Evaluating this arrangement
requires accounting for both retained information and the width of each
retrieval response. Decision-tree capacity bounds describe individual
sets of distinguishable records, but a memory writer must encode every
possible record before the query is known. We study fixed-length retention
beside an immutable binary archive, with a worst-case limit on raw bit or
word reads. We characterize feasible one-bit-read memory cells and
construct encoders that partition the complete record space. Supporting
pairwise exclusive disjunction and conjunction with one bit read is
equivalent to supporting every Boolean query. For multiple reads, we
combine explicit encoders with decision-tree counting and an entropy
converse. The word-read bounds differ by at most two retained bits when
the read budget is at most half the archive length. Under an explicit
condition on word width and archive length, raw-word caching attains the
exact optimum among all writers. A related condition preserves this
optimum under positive average error. Exhaustive checks validate the implementations.
In a matched dialogue-agent study, a categorical sufficient statistic
preserves source decisions with less storage and lower decoding latency
than the universal encoder. A catalogue-specific index reduces private
state further and requires additional shared storage.
The results characterize how response width and future query structure
determine retention requirements.
\end{abstract}
\maketitle
\hypersetup{pdfauthor={Pranav Singh and Prashant Singh}}

\section{Introduction}
An agent may receive evidence before knowing which future decision will
use it. A memory writer can compress that evidence into retained state,
while a later reader retrieves selected parts of the original archive.
The retained state and retrieval responses jointly determine which
decisions remain possible. Decision-oriented compression and selective
information acquisition already address this relationship
\parencite{permuter2011,bilgic2011,zou2026}.

A useful comparison must specify the archive interface. A bit read
returns one original binary coordinate. A word read returns a fixed
block of original coordinates. Both interfaces differ from querying a
rewritten index or a semantic retrieval service. Systematic data
structures formalize storage beside original data
\parencite{gal2003,ramamoorthy2020}. Input-dependent information available
without charged probes also appears as published bits \parencite{patrascu2010}.
Help-bit models study information
provided before bounded decision-tree computation \parencite{nisan1999}.
Our focus is a single writer shared by all subsequent Boolean queries,
where a Boolean query returns a binary decision about the record.

Decision-tree capacity provides a local restriction on such a writer.
A retained state identifies a set of possible records, called its memory
cell. The reader must answer each query on that cell using bounded raw
retrieval. Established shattering results characterize sets on which
decision trees can realize every binary labeling
\parencite{yildiz2012,yildiz2015,leboeuf2022}.
A retention guarantee also requires a fixed-length encoding that
partitions the complete record space into suitable cells.
Response width must be included when interpreting the resulting savings.

\begin{figure}[t]
\centering\includegraphics[width=\linewidth]{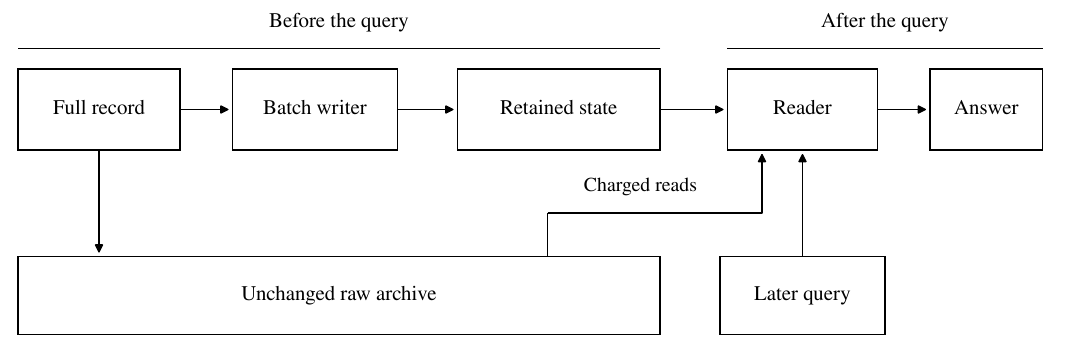}
\caption{Pre-query writing and bounded raw retrieval.}
\label{fig:pipeline}
\Description{A full record enters the writer before a query exists.
The writer produces fixed-length retained state. Later the reader uses
that state, a query, and charged accesses to the unchanged raw archive
to produce a Boolean answer.}
\end{figure}

We study this global encoding problem with unrestricted computation
and an explicit worst-case read budget. Figure~\ref{fig:pipeline} shows
the writer and reader interfaces. The writer records offsets relative
to public anchor features. After receiving a query, the reader evaluates
it on the candidate records allowed by those offsets. A selected raw
coordinate then determines the answer. For word retrieval, the reader
compares this construction with retaining all words except those it
will read directly.

A retained payload alone does not specify archive requirements or query
coverage. The matched application compares these resources under fixed
source observations and includes shared codebooks in its storage accounting.

Our contributions are as follows:
\begin{itemize}
\item We characterize one-bit-read memory cells and prove that supporting
pairwise exclusive disjunction and conjunction implies support for every
Boolean query. Explicit full-domain encoders attain the resulting
fixed-bit retention bounds.
\item We derive word-read retention bounds with a uniform gap of at most
two bits for budgets up to half the archive length. We identify a regime
in which raw-word caching is exactly optimal among all writers.
\item We extend the counting converse to randomized schemes with average
error and obtain an exact caching regime with positive error allowance.
\item We provide instrumented implementations, exhaustive checks, and a
matched text-to-decision evaluation. The study separates preservation of
source decisions from source accuracy and compares payload, archive,
and shared storage across query workloads.
\end{itemize}
The capacity bounds and selector constructions originate in prior
decision-tree work \parencite{yildiz2012,yildiz2015}. Our results concern
writer partitions, workload equivalence, and their retained-state
consequences under specified retrieval interfaces.

\section{Related work}\label{sec:related}
\paragraph{Systematic storage and bit probes.}
\textcite{gal2003} study original data with an auxiliary index for prefix
sums and substring search. Their accounting charges probes into the
stored structure. Our reader sees the entire retained tier directly.
\textcite{viola2012} proves redundancy lower bounds for encoded ternary arrays
and set membership under adaptive bit probes.
\textcite{patrascu2010} establish word-probe bounds for rank and select.
Their analysis uses input-dependent published bits that the reader
examines without charge. Our retained tier has this established role.
We study the stated pair and universal workloads while restricting
charged access to the unchanged record.

\paragraph{Help bits and linear structures.}
\textcite{nisan1999} study help bits followed by decision trees that
evaluate copies of a fixed function on disjoint inputs. Our writer
supports future queries that may overlap on the same record.
\textcite{ramamoorthy2020} relate systematic linear structures for inner
products to matrix rigidity. We allow nonlinear writers and analyze
universal Boolean queries, including a comparison with affine writers.

\paragraph{Decision-tree capacity.}
\textcite{yildiz2012} establish the capacity of binary and multiway stumps.
They also provide a selector construction. A stump is a decision tree
with one internal test. The journal treatment extends these results
\parencite{yildiz2015}. \textcite{leboeuf2022} analyze partitioning and
generalization for real-valued and categorical features.
We use the binary-stump and selector ingredients to construct global
writer partitions. Multiway tree counting yields our word-width bounds.
We attribute these capacity tools to their predecessors.

\paragraph{Information acquisition and decision compression.}
Side-information vending jointly designs a compressed description and
later information-acquisition actions \parencite{permuter2011}.
Locally decodable source coding limits access to a coded description
\parencite{makhdoumi2013}. Our reader accesses original archive coordinates.
Value-of-information methods select features using probabilistic
structure \parencite{bilgic2011}. Approximate information states preserve
information required for planning \parencite{subramanian2022}.
Our universal-query objective provides a worst-case comparison for
retention beside a raw archive.

\paragraph{Agent memory.}
Finite recurrent-state retrieval limits \parencite{wen2024} and
rate--distortion formulations for agent memory \parencite{zou2026}
motivate explicit state accounting. Nous studies belief maintenance
and certification \parencite{singh2026nous,singh2026certification}.
We reuse the pretrained models and the released updater \parencite{nousrelease}
in a matched text-to-decision evaluation. The observation stage is held
fixed within each source group while the retained representation changes.

\subsection{Retained-state contribution}
A capacity bound concerns the labelings realizable on one set of records.
Our writer must partition the complete input space before the query is
known. Table~\ref{tab:novelty} distinguishes the imported ingredients from
the retained-state statements established in this work.
The universal one-read converse is a fixed-length consequence of the
known stump capacity. Its attaining encoder requires a compatible
global partition. The multiple-read construction lifts the established
selector construction through public anchors and stored offsets.
The entropy and word results extend counting to the stated retention
and error conventions. Global preprocessing is standard in data structures
\parencite{gal2003,viola2012}. Our contribution is the stated workload
characterization and retained-state bounds within that framework.

\begin{table}[t]
\centering\small\setlength{\tabcolsep}{4pt}
\begin{tabular}{p{.15\linewidth}p{.37\linewidth}p{.40\linewidth}}\toprule
Result & Established ingredient & Retained-state conclusion\\\midrule
Theorem~\ref{thm:structure} & Coordinate stumps and binary shattering capacity \parencite{yildiz2012,yildiz2015} & Affine-hull characterization and equivalence of pairwise XOR-plus-AND support with universal one-read support.\\
Theorem~\ref{thm:exact} & Binary-stump capacity and simplex-code features & Full-domain offset partitions attaining fixed-bit optima, with an affine-writer comparison.\\
Theorem~\ref{thm:multi} & Selector shattering \parencite{yildiz2012} and decision-tree counting \parencite{leboeuf2022} & Compatible global encoders and converse bounds, with an exact affine-writer optimum.\\
Theorem~\ref{thm:approx} & Counting finite tree descriptions and the binary entropy inequality & A converse for randomized writers fixed before the query and uniform-source average error.\\
Theorem~\ref{thm:words} & Multiway capacity \parencite{yildiz2012,yildiz2015} and multiway tree counting & Word-width bounds, a uniform two-bit gap, and an exact raw-caching regime.\\
Proposition~\ref{prop:word-error} & The entropy converse and the word-tree count & A finite-error condition under which integer retention still makes raw caching optimal.\\\bottomrule
\end{tabular}

\caption{Established ingredients and retained-state conclusions.
XOR denotes exclusive disjunction and AND denotes conjunction.}
\label{tab:novelty}
\end{table}

\section{Model and resource conventions}\label{sec:model}
A record $x\in\bits^n$ is available in full during batch writing.
A public writer $W$ emits a fixed state $m=W(x)\in\bits^B$ before the
actual query is supplied. The archive remains the original indexed
array $(x_1,\ldots,x_n)$. The reader sees the state and query, and requests
at most $R$ original coordinates. Each response contains one bit.
Later addresses may depend on earlier responses.
Section~\ref{sec:words} extends the interface to aligned words.

We consider three query families: exclusive disjunction (XOR) of every
distinct coordinate pair, every Boolean gate on each pair, and every
Boolean function of the complete record. Conjunction is denoted AND,
and disjunction is denoted OR. A query is a pure evaluable function of
a hypothetical record. A truth table is an allowed finite description.
The writer may know the query family and record length, but does not
know the eventual query. Exact guarantees hold for every record and query.

Public code and schema are independent of the record. All persistent
record-dependent indexes, parameters, seeds, and metadata belong in the
retained state. Each query starts from that state without a reusable
transcript of earlier queries. Random tapes are independent of the
record and query. The read limit holds on every execution.
The raw archive and retained tier together occupy $n+B$ logical bits.
The tier is fully visible, while only archive probes count toward $R$.
Charging every retained-bit access would allow simulation with $B+R$ bit probes
by reading the entire tier first.
An affine writer has the form $W(x)=Ax+b$ over the binary field,
with a fixed matrix $A$ and vector $b$.
The theory allows unrestricted computation and transient workspace.
Writing is a batch operation with access to the complete record.

The bit implementation stores each binary fact in one byte and packs
the retained state into $\lceil B/8\rceil$ bytes with zero padding.
The word implementation returns the complete declared word value.
Archive storage, transient allocations, shared code, and extraction
models are accounted for separately from retained bits.

\section{One-read cells and exact retention}\label{sec:theory}
The key object is the set of histories that the retained state fails to
distinguish. For a state $m$, let $S_m=W^{-1}(m)$, omitting empty cells.
On a cell $S$, write $c_i(x)=x_i$ and define
\begin{equation}
 \mathcal D(S)=\{0,1,c_1,1+c_1,\ldots,c_n,1+c_n\}.
\end{equation}
All arithmetic in this section is over $\mathbb F_2$. A reader that knows
$m$ can select an address separately for each query. Once that address is
selected, a function of the returned bit is constant, identity, or
complement. Consequently, a query is exactly one-read answerable on $S$
if and only if its restriction belongs to $\mathcal D(S)$.
Different queries may select different addresses from the retained state.

\begin{theorem}[Structure of one-read cells]\label{thm:structure}
Fix a nonempty $S\subseteq\bits^n$, a reference $x^0\in S$, and
$V=\operatorname{span}\{x+x^0:x\in S\}$. Let $d=\dim V$, and let $V^*$ denote its dual space of linear forms.
\begin{enumerate}
\item Every pairwise XOR is one-read answerable on $S$ if and only if
the coordinate linear forms restricted to $V$, together with zero,
exhaust $V^*$. In this case $d\le\lfloor\log_2(n+1)\rfloor$, and the
entire affine hull $x^0+V$ is also XOR-feasible.
\item The following conditions are equivalent:
\begin{enumerate}
\item All Boolean functions of each pair are one-read answerable.
\item All pairwise XOR and AND functions are one-read answerable.
\item Every Boolean function on $S$ is one-read answerable.
\item The cell $S$ is XOR-feasible and affinely independent.
\end{enumerate}
\end{enumerate}
\end{theorem}

The first part requires the raw coordinates to contain every linear
measurement direction on the remaining uncertainty space. The second part requires affine independence when conjunctions and
parity relations are both supported. The same writer cells then support arbitrary Boolean queries. Appendix~\ref{app:structure} proves the theorem.

\begin{theorem}[Exact fixed-bit optima]\label{thm:exact}
For $n\ge2$, with at most one raw bit read and exact answers on the full
cube, the minimum retained state lengths are
\begin{align}
 B_{\oplus}(n,1)&=n-\lfloor\log_2(n+1)\rfloor,\label{eq:xor}\\
 B_{\mathrm{pair}}(n,1)=B_{\mathrm{all}}(n,1)
 &=n-\left\lfloor\log_2\!\left(1+\lfloor\log_2(n+1)\rfloor\right)\right\rfloor .
 \label{eq:universal}
\end{align}
The universal optimum can be attained by a nonlinear writer. If the
writer is restricted to fixed affine measurements over $\mathbb F_2$,
the optimum for the Boolean-pair workload is $n-1$.
\end{theorem}

The retained length $B$ is an integer number of fixed bits.
For the XOR workload, every feasible cell has at most
$2^{\lfloor\log_2(n+1)\rfloor}$ points. For the universal workload it has
at most $s_n=1+\lfloor\log_2(n+1)\rfloor$ points.
The latter formula is the known Vapnik--Chervonenkis (VC) dimension
of binary decision stumps \parencite{yildiz2012,yildiz2015}. 
The retention problem also requires a partition of every history
into decodable cells. The constructions below supply that global partition,
without a history-specific codebook.
The construction uses power-of-two-sized cells and attains the
fixed-bit optimum.

\subsection{An attaining XOR encoder}\label{sec:xor-code}
Set $r=\lfloor\log_2(n+1)\rfloor$. Assign public labels
$v_1,\ldots,v_n\in\mathbb F_2^r$ containing all $2^r-1$ nonzero vectors,
and use zero labels for any remaining positions. Choose the positions of
the $r$ standard basis vectors as anchors. Let $z$ be the corresponding
raw archive bits. The writer retains the $n-r$ nonanchor offsets
\begin{equation}
 a_i=x_i+\langle v_i,z\rangle.
\end{equation}
Anchor offsets are zero. For query $(i,j)$,
if $v_i+v_j=0$, return $a_i+a_j$. Otherwise choose the public position
$\ell$ labeled $v_i+v_j$, read $x_\ell$, and return
$x_\ell+a_\ell+a_i+a_j$. The returned value equals $x_i+x_j$.
The cells are cosets of a binary simplex-code subspace with zero padding.
The construction uses this standard code structure.

\subsection{An attaining universal encoder}\label{sec:bool-code}
Set $k=\lfloor\log_2 s_n\rfloor$ and $t=2^k$. List as public features
$\phi_1,\ldots,\phi_n:\bits^k\to\bits$ every nonzero Boolean function
satisfying $\phi(0)=0$, then pad with zero functions. The feature list fits
because there are $2^{t-1}-1\le n$ nonzero normalized functions.
The list includes the $k$ identity functions. Use their positions as
anchors, let $z$ be the corresponding raw bits, and retain
\begin{equation}
 a_i=x_i+\phi_i(z)
\end{equation}
at the $n-k$ nonanchor positions. All anchor offsets are zero.
Every payload defines exactly the $t$ candidate histories
$x(z)=a+\Phi(z)$, where $\Phi(z)=(\phi_1(z),\ldots,\phi_n(z))$.

Given any pure Boolean query $q$, compute its truth table on these
candidates, $g(z)=q(a+\Phi(z))$. Put $b=g(0)$.
If $g+b$ is zero, answer $b$ without reading. Otherwise the public feature
list contains $g+b=\phi_\ell$. Read $x_\ell$ and answer $x_\ell+a_\ell+b$.
The query evaluation uses hypothetical histories obtained from the payload. The read address is fixed before the
read response arrives.

This procedure makes $t$ evaluations of $q$. For a pair query it can
construct $g$ directly from two feature tables and the two offsets,
without materializing candidate histories. The exhaustive-label ingredients are
established in stump shattering constructions \parencite{yildiz2012}.
The offset encoder turns them into a full-cube partition.

\paragraph{Workload dependence.}
With two raw reads, either pair workload needs zero retained state.
With no reads, pairwise XOR needs $n-1$ bits and the Boolean-pair workload
needs $n$. The former identifies a record up to global complementation.
The latter includes every coordinate projection.
Appendix~\ref{app:endpoints} proves these endpoint statements.
\section{Universal queries with multiple reads}\label{sec:multi}
Let $B_{\mathrm{all}}(n,R)$ denote the optimum for all Boolean queries
under the same full-cube, fixed-state model. This workload is larger
than pair queries when $R\ge2$.

\begin{theorem}[Uniform bounds for universal retrieval]\label{thm:multi}
For $n\ge2$ and $1\le R\le n$, define
\begin{equation}
 C_{n,R}=2^R+(2^R-1)\log_2 n,\qquad
 k_{n,R}=\left\lfloor\log_2\!\left(1+
                   \lfloor\log_2(n-R+2)\rfloor\right)\right\rfloor .
\end{equation}
Then
\begin{equation}
 \max\{0,n-\lfloor\log_2\lfloor C_{n,R}\rfloor\rfloor\}
 \le B_{\mathrm{all}}(n,R)
 \le n-R+1-k_{n,R}.\label{eq:multi}
\end{equation}
In particular, uniformly for $1\le R\le\lfloor n/2\rfloor$,
\begin{equation}
 n-R-\log_2\log_2 n-1
 \le B_{\mathrm{all}}(n,R)
 \le n-R-\log_2\log_2 n+3.\label{eq:uniform}
\end{equation}
At $R=0$ the optimum is $n$, and at $R=n$ it is zero.
For affine writers the optimum for arbitrary Boolean queries is exactly
$n-R$ for $0\le R\le n$.
\end{theorem}

The converse is a standard decision-tree counting argument applied inside
each memory cell. There are at most
$n^{2^R-1}2^{2^R}$ full depth-$R$ trees with raw-coordinate tests and
Boolean leaves. Universal answerability requires these trees to
realize every labeling of the cell. Decision-tree counting and shattering
are established tools \parencite{yildiz2012,leboeuf2022}.

For the construction, allocate $R-1$ coordinates as selector anchors and
apply the universal feature construction to the remaining $n-R+1$
coordinates, with $k_{n,R}$ tail anchors. Retain only the offsets on all
nonanchors. The reader first reads the selector bits, then evaluates
the query on the $2^{k_{n,R}}$ possible tail histories and performs the
single tail-feature read. Adaptivity is used between the selector
stage and the final read. The degenerate selector-plus-stump construction
appears in \textcite[Theorem 2]{yildiz2012}. Our full binary selector stage
corresponds to the construction in \textcite[Theorem 3]{yildiz2012}.
The offset encoding partitions the entire raw-history cube into decodable cells.
Appendix~\ref{app:multi} gives the partition and rounding details.

Equation~\eqref{eq:uniform} limits universal retained-state savings.
Beyond the $R$ bits recoverable by storing everything else, one can omit
only about $\log_2\log_2 n$ further bits, up to constants. The affine-writer
bound shows an unbounded additive nonlinear advantage, with no
constant-factor reduction in retained state when $R=o(n)$.
A compact semantic statistic can be much smaller for restricted history
support or a restricted future query family.

\section{Average-error guarantees}\label{sec:approx}
Worst-case exactness is a strong requirement. A related obstruction
survives if each query may err on a fixed fraction of uniformly sampled
records. Let $B_\epsilon(n,R)$ be the minimum fixed retained length such
that, for every Boolean query $q$,
\begin{equation}
 \Pr_{X,\Omega}\{\widehat q(X)\ne q(X)\}\le\epsilon,\qquad
 X\sim\operatorname{Unif}(\bits^n),
\end{equation}
where $\Omega$ denotes all algorithmic randomness independent of $X$
and the query. The read limit holds in every execution. Abstention is
counted as an error. A query may be selected adversarially but is
fixed independently of the particular drawn record.

For the zero-read case, set $C_{n,0}=1$.

\begin{theorem}[Average-error universal converse]\label{thm:approx}
For $n\ge2$, $0\le R\le n$, and $0\le\epsilon<1/2$,
\begin{equation}
 B_\epsilon(n,R)\ \ge\
 \max\left\{0,\left\lceil n-\log_2 C_{n,R}
                  +\log_2(1-h_2(\epsilon))\right\rceil\right\},
\end{equation}
where $h_2(p)=-p\log_2 p-(1-p)\log_2(1-p)$ is binary entropy,
with $0\log_2 0=0$. Shared and private independent randomization
are permitted. For $R\ge1$, the exact construction gives the upper bound in
Equation~\eqref{eq:multi}. For $R=0$, retaining all $n$ bits suffices.
Consequently, for each fixed
$\epsilon<1/2$, uniformly over $1\le R\le\lfloor n/2\rfloor$,
\begin{equation}
 B_\epsilon(n,R)=n-R-\log_2\log_2 n+O_\epsilon(1).
\end{equation}
\end{theorem}

The proof chooses a random Boolean query, whose labels on the $2^n$
records are independent fair bits. Conditioned on the random tapes, each
writer cell must approximate these labels using one of at most
$2^{C_{n,R}}$ decision-tree output patterns. A binary entropy argument applied to the vector of
state-indexed trees yields the bound, including randomized writers. Appendix~\ref{app:approx} proves every step.
The entropy inequality is a standard information-theoretic tool. The
claim concerns its application to a writer fixed before the query.
This quantifier matters. If a separate writer may be chosen after seeing
$q$, it can retain the one-bit answer $q(x)$ and avoid retrieval.

\section{Universal queries with word retrieval}\label{sec:words}
Physical reads commonly return several bits together. We therefore extend
the archive interface to fixed, disjoint words. Let $M\ge1$ be the number
of words and $w\ge1$ their width in bits. The full record has $N=Mw$ bits,
and each read returns one of the $M$ original words without modification.
The response alphabet has size $Q=2^w$. The writer retains $B$ bits before
receiving the query. Define $B^{(w)}_{\mathrm{all}}(M,R)$ as the minimum
retention for every Boolean query, with at most $R$ adaptive word reads.

\begin{theorem}[Word-width bounds and exact caching regime]\label{thm:words}
For $1\le R\le M$, define
\begin{align}
 C^{(w)}_{M,R}&=Q^R+\frac{Q^R-1}{Q-1}\log_2 M,\label{eq:word-count}\\
 k^{(w)}_{M,R}&=\left\lfloor\log_2\!\left(1+
 \lfloor\log_2(w(M-R+1)+1)\rfloor\right)\right\rfloor.
 \label{eq:word-k}
\end{align}
The minimum retained length satisfies
\begin{align}
 \max\{0,N-\lfloor\log_2\lfloor C^{(w)}_{M,R}\rfloor\rfloor\}
 &\le B^{(w)}_{\mathrm{all}}(M,R),\label{eq:word-lower}\\
 B^{(w)}_{\mathrm{all}}(M,R)
 &\le N-w(R-1)-\max\{w,k^{(w)}_{M,R}\}.\label{eq:word-upper}
\end{align}
If $\log_2 M\le 2^w-1$, then, for every $0\le R\le M$,
\begin{equation}
 B^{(w)}_{\mathrm{all}}(M,R)=N-wR.\label{eq:word-exact}
\end{equation}
For $M\ge2$ and $1\le R\le\lfloor M/2\rfloor$, the upper and lower
bounds in Equations~\eqref{eq:word-lower}--\eqref{eq:word-upper} differ
by at most two bits. Their constants are independent of the word width.
\end{theorem}

The proof counts coordinate decision trees with a branch for each word
value. Multiway decision-tree capacity is established in
\textcite[Theorem 5]{yildiz2012} and \textcite[Theorem 3]{yildiz2015}.
Theorem~\ref{thm:words} applies this counting argument to a complete
writer partition and supplies an attaining cache in the stated regime.
Appendix~\ref{app:words} gives the proof and the uniform rounding argument.

The constructive bound compares two writers. A raw-word cache retains
all words except a public set of $R$ words. Reading those words
reconstructs the complete record. The alternative reads $R-1$ selector
words and applies the universal bit encoder to the remaining bits.
Its final bit probe is implemented by reading the containing aligned
word and selecting the required bit. The writer chooses whichever
construction retains fewer bits, using only the public dimensions.

Equation~\eqref{eq:word-exact} makes response width consequential for
universal retention. For eight-bit words, the equality holds whenever
$M\le2^{255}$. A nonlinear writer cannot reduce the fixed retained
length below raw caching in this regime. The statement concerns
logical retained bits, with the full raw archive still present.
The equality also covers adaptive readers and zero-error randomized schemes.

For the zero-read case, set $C^{(w)}_{M,0}=1$.

\begin{proposition}[Word retrieval with average error]\label{prop:word-error}
For $M,w\ge1$ and $0\le R\le M$, let $B^{(w)}_\epsilon(M,R)$ allow error at most $\epsilon<1/2$ for each
fixed Boolean query under the uniform record distribution and independent
random tapes. Every execution uses at most $R$ word reads. Then
\begin{equation}
 B^{(w)}_\epsilon(M,R)\ge
 \max\left\{0,\left\lceil N-\log_2 C^{(w)}_{M,R}
       +\log_2(1-h_2(\epsilon))\right\rceil\right\}.
 \label{eq:word-error}
\end{equation}
For $1\le R\le M$, if
\begin{equation}
 1+(1-Q^{-R})\frac{\log_2 M}{Q-1}
       <2(1-h_2(\epsilon)),\label{eq:word-error-regime}
\end{equation}
then $B^{(w)}_\epsilon(M,R)=N-wR$.
A sufficient condition is $h_2(\epsilon)<1/2$ and
$\log_2 M\le(Q-1)(1-2h_2(\epsilon))$.
\end{proposition}

The entropy argument from Theorem~\ref{thm:approx} applies to the
multiway tree count. Integer rounding then establishes an exact
optimality regime even with positive average error.
The sufficient condition quantifies how source uncertainty, response
width, and allowed error interact. The condition imposes the same guarantee for
every query and does not assume that the query distribution is known.

\section{Implementation and evaluation}\label{sec:implementation}
The implementations return a fixed-length byte payload from the writer.
The reader receives only that payload, a pure query evaluator, and an
instrumented archive callback. The callback records every address and
rejects budget overruns. Public codec objects contain dimensions and
feature rules. The reader validates payload length and padding.

The generic bit decoder evaluates a query on $2^{k_{n,R}}$ candidate
records. Its direct implementation uses transient storage for offsets
and candidate records. A specialized pair-gate decoder evaluates feature
truth tables without materializing complete candidates.
The word decoder reads selector words and obtains its final selected
bit from the containing word. When raw caching retains fewer bits,
the implementation instead reconstructs the record from cached and
retrieved words.

We compare the universal encoder with a matched raw-prefix cache.
The cache enumerates its compatible records and selects the coordinate
that maximizes answerable mass under the uniform source.
A branch is answerable when every compatible record gives the same
query value. The cache abstains on other branches. We also compare
complete-record retention, the exact XOR encoder, and a categorical
sufficient statistic for the dialogue-derived records.
Appendix~\ref{app:reproduction} specifies sampling and implementation details.

\subsection{Computational checks}\label{sec:experiments}
\begin{figure}[t]
\centering\includegraphics[width=\linewidth]{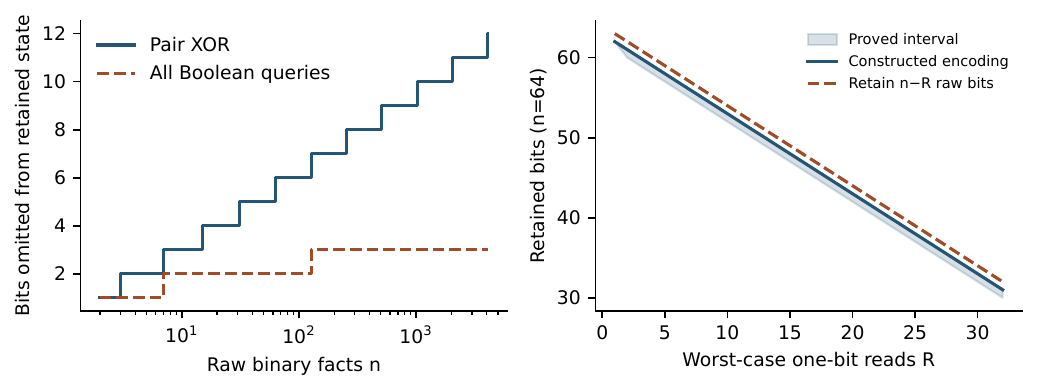}
\caption{Retention bounds for bit retrieval.}
\label{fig:frontiers}
\Description{The left panel compares one-read savings by query family.
The right panel compares universal bounds and affine retention as the
read budget increases.}
\end{figure}

\begin{figure}[t]
\centering\includegraphics[width=\linewidth]{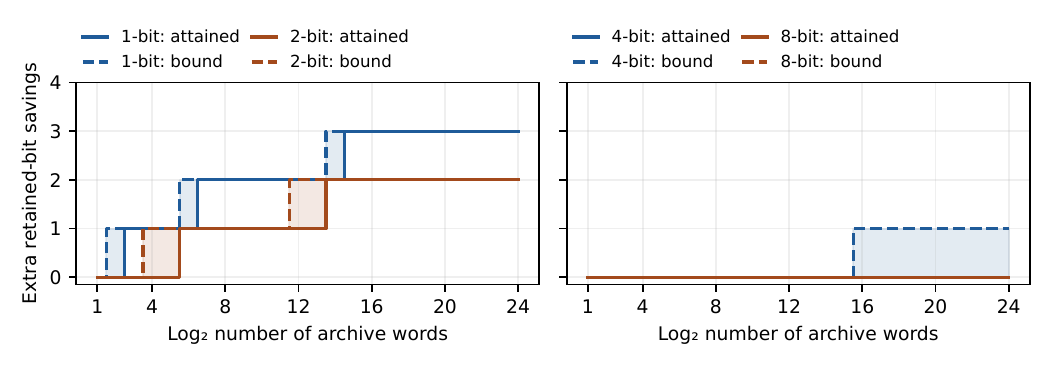}
\caption{One-word-read savings and upper bounds relative to raw caching.}
\label{fig:width}
\Description{Bounds on the extra savings decrease as the response width
increases. The exact caching regime has zero extra savings.}
\end{figure}

Table~\ref{tab:checks} summarizes the finite checks. The construction
checks enumerate records and queries, compare decoded answers with
direct evaluation, and enforce the read budget.
Independent cell enumeration compares the structural characterization
with attainable truth tables. The word check also solves minimum
partitions of small cubes by direct subset enumeration.
All reported checks passed. Figure~\ref{fig:frontiers} shows the proved
bit-read bounds, and Figure~\ref{fig:width} shows the word-width transition.

\begin{table}[t]
\centering\small\begin{tabular}{lr}\toprule
Check & Cases\\\midrule
XOR construction & 458,748\\
Boolean-pair construction & 1,212,352\\
Arbitrary truth tables & 6,272\\
Multi-read construction & 699,608\\
Independent binary-cell enumeration & 65,808\\
Independent codec checks & 201,660\\
Word-codec truth tables & 12,624\\
Word-codec parity and point queries & 716,800\\
Word-codec sampled threshold queries & 800\\
\bottomrule\end{tabular}

\caption{Construction and structural checks. XOR denotes exclusive
 disjunction.}
\label{tab:checks}
\end{table}

\subsection{Query shift and archive failures}
We evaluated pair XOR, pair AND, masked parity, and masked threshold
queries under uniform independent records. Each comparison used the
same query and record for both methods. The writer supported every
query family without refitting. Table~\ref{tab:controlled-protocol} gives the
sample counts and conditions. We measured coverage, error among answered
queries, and mean reads. Wilson intervals in the accompanying results
summarize coverage and conditional error.

\begin{figure}[t]
\centering\includegraphics[width=\linewidth]{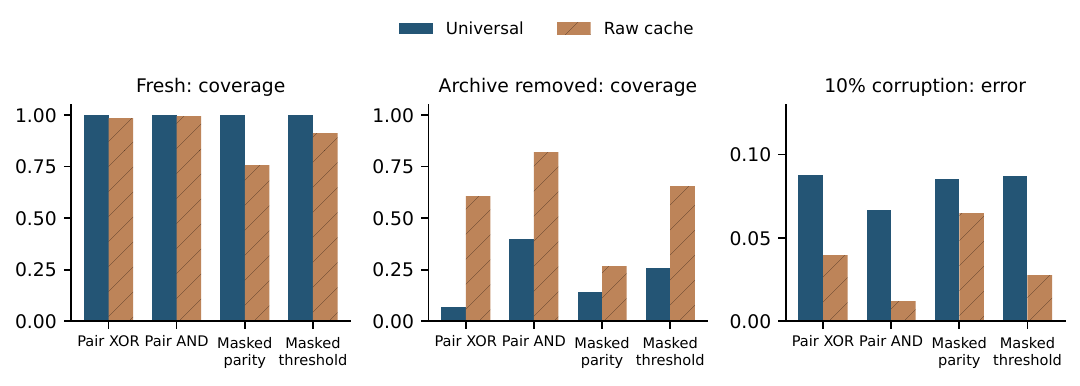}
\caption{Coverage and conditional error under archive failures.}
\label{fig:failures}
\Description{The panels compare fresh coverage, removed-archive coverage,
and conditional error under archive corruption for matched retained budgets.}
\end{figure}

The stale-archive condition independently corrupted archive bits after
writing. Ground truth remained the query value on the original record.
The removed-archive condition returned abstention whenever a read was
required. Table~\ref{tab:controlled} reports the smallest record-size
condition. The masked-query rows show lower fresh coverage for the raw
cache, which often used fewer reads. The removed-coverage column favors
the raw cache for several query families. Both methods made errors under
archive corruption. Figure~\ref{fig:failures} compares these conditions.

\begin{table}[t]
\centering\small
\begin{tabular}{ll}\toprule
Quantity & Value\\\midrule
Record lengths (bits) & 10, 16, 64\\
Worlds per method-condition cell & 2,000\\
Method-condition cells & 96\\
Archive bit-corruption probabilities & 0.01, 0.10\\
Displayed condition & $n=10$, $B=8$, $R\le1$, corruption $0.10$\\
\bottomrule\end{tabular}
\caption{Controlled-source evaluation protocol.}
\label{tab:controlled-protocol}
\end{table}

\begin{table}[t]
\centering\small\begin{tabular}{llrrrr}\toprule
Query & Method & Coverage & Reads & Removed coverage & Stale error\\\midrule
Pair XOR & Universal & 1.000 & 0.937 & 0.068 & 0.088\\
Pair XOR & Raw cache & 0.983 & 0.346 & 0.607 & 0.040\\
Pair AND & Universal & 1.000 & 0.612 & 0.401 & 0.067\\
Pair AND & Raw cache & 0.992 & 0.193 & 0.819 & 0.012\\
Masked parity & Universal & 1.000 & 0.875 & 0.141 & 0.086\\
Masked parity & Raw cache & 0.754 & 0.506 & 0.267 & 0.065\\
Masked threshold & Universal & 1.000 & 0.730 & 0.261 & 0.087\\
Masked threshold & Raw cache & 0.910 & 0.359 & 0.653 & 0.027\\
\bottomrule\end{tabular}

\caption{Coverage, mean bit reads, and conditional stale error.
XOR denotes exclusive disjunction. AND denotes conjunction.}
\label{tab:controlled}
\end{table}

\subsection{Matched dialogue-agent evaluation}\label{sec:agent}
We evaluated retention in a text-to-decision pipeline for dialogue-based
offer validation. The writer processes a conversation before learning
which catalogue entity the agent will check. A query specifies an entity
and domain. The reader decides whether the entity satisfies the tracked
categorical constraints. The task uses the public MultiWOZ catalogues
and annotated dialogue states \parencite{multiwoz24}.

We used two fixed observation sources from the original pretrained Nous
models, provided with the package. The source code accompanies the Nous
release \parencite{nousrelease}. The native source classifies each user
utterance with its preceding system turn, then applies the native Nous
updater. The history source classifies the complete observed prefix
at compaction. We performed fresh inference without fitting either
source. Every retention method receives identical observations within
each source group. We compare retention methods within those groups.

The primary query population contains goal-active domains with at least
one annotated constraint among the tracked slots. We tested every
catalogue entity in each eligible domain. Gold labels determine the
evaluation population and correctness, and never enter the writer.
An all-domain query population tests sensitivity to the requested
workload. Table~\ref{tab:agent-protocol} gives the dataset and query counts.

The posterior summary stores every source probability at double precision.
The category vector stores each selected category separately.
The mixed-radix representation stores the complete categorical answer
state. The universal bit encoder uses the corresponding binary indicators
and obtains its selected bit from an aligned byte.
Raw caches retain a public suffix of that packed record.
Their readers inspect uncached words only when needed to decide the query.
The raw readers also exploit the same public categorical support for
short-circuiting. All archive methods use the same packed record and
charge each complete byte response.
The theoretical bounds fix a worst-case read budget.
The dialogue study compares policies with different read limits.
Observation streams and query populations are shared within each source group.
The universal encoder permits one byte read per query.
For an archive of $M$ words, retaining $c$ cached words permits $M-c$ reads.
State-only methods permit no reads.
Table~\ref{tab:agent-native} records these limits beside mean returned bytes.

\begin{table}[t]
\centering\small\begin{tabular}{ll}\toprule
Quantity & Value\\\midrule
Official test dialogues & 1,000\\
Included final states & 999\\
Observed user turns & 7,372\\
Primary dialogue clusters & 738\\
Primary queries per source & 67,346\\
Compatible / incompatible primary queries & 19,242 / 48,104\\
Catalogue sizes: hotel / restaurant / attraction & 33 / 110 / 79\\
All-domain queries per source & 221,778\\
Original methods / added catalogue baseline & 9 / 1\\
Archive conditions & 3\\
Bootstrap replicates & 2,000\\
Packed raw archive (B/state) & 4\\
Shared model bundle (B) & 29,791,103\\
Shared dialogue archive (B) & 13,790,771\\
Catalogue classes: hotel / restaurant / attraction & 41 / 24 / 6\\
Combined catalogue-answer classes & 5,904\\
Catalogue minimum bits / payload bytes & 13 / 2\\
Additional public codebook (B) & 8,878\\
\bottomrule\end{tabular}

\caption{Matched dialogue-agent evaluation protocol. B denotes bytes.}
\label{tab:agent-protocol}
\end{table}

A catalogue-specific baseline compiles the public answer-equivalence
classes of the categorical support. Let $\mathcal C_d$ contain the
distinct catalogue-answer patterns for domain $d$.
Domain-factorized states realize every combination of these patterns.
An exact zero-read writer therefore requires
\begin{equation}
 B_{\mathrm{cat}}=\left\lceil\log_2\prod_d|\mathcal C_d|\right\rceil.
 \label{eq:catalogue}
\end{equation}
A mixed-radix index of the domain patterns attains this bound.
The public codebook is fixed by the catalogue and ontology.
Its serialized size is charged separately from the episodic payload.
Table~\ref{tab:agent-storage} includes that cost in cohort storage.
The construction is a finite-query sufficient-statistic baseline.

Table~\ref{tab:agent-quality} reports correctness against annotated
constraints and balanced accuracy. Every fresh retained-state method
preserves its source decisions. The history source has higher correctness
than the native update source. The rejection control exposes the class
imbalance in the primary population. Its raw accuracy exceeds that of
the native source, while the native source has higher balanced accuracy.
Within each source group, fresh encoding leaves decision accuracy unchanged.
These measures separate source-decision quality from memory preservation.

\begin{table}[t]
\centering\small\begin{tabular}{lrrr}\toprule
Decision source & Accuracy & Balanced accuracy & 95\% interval\\\midrule
Native update & 0.4725 & 0.6170 & [0.4487, 0.4978]\\
History classifier & 0.8070 & 0.8461 & [0.7846, 0.8288]\\
Always reject & 0.7143 & 0.5000 & [0.6992, 0.7293]\\
Always accept & 0.2857 & 0.5000 & [0.2707, 0.3008]\\
\bottomrule\end{tabular}

\caption{Offer-validation accuracy and dialogue-bootstrap intervals.
Intervals concern accuracy.}
\label{tab:agent-quality}
\end{table}

Table~\ref{tab:agent-native} reports the primary resource comparison.
The mixed-radix state preserves the posterior summary's decisions with
a smaller payload and no archive reads. The catalogue index reduces
the payload further by preserving only the fixed tool-query family.
Its extra codebook increases cohort storage at the studied scale.
The arithmetic mixed-radix representation uses less aggregate storage.
In the query-latency column, mixed radix and the catalogue statistic
are faster than the universal bit encoder.
The raw-cache row sharing the universal encoder's cap uses less retained
state and has lower median query latency.
The universal encoder retains its archive and evaluates candidate records.
Its guarantee covers the complete binary record space.
The categorical representation preserves the complete extracted category state.
The catalogue index preserves its fixed family of entity decisions.
These guarantees use different record supports and query families.

\begin{table}[t]
\centering\scriptsize\begin{tabular}{lrrrrrrr}\toprule
Method & State (B) & Archive (B) & $R_{\max}$ & Returned (B/query) & Write ($\mu$s) & Query ($\mu$s) & p95 ($\mu$s)\\\midrule
Posterior summary & 312 & 0 & 0 & 0.0000 & 31.042 & 8.250 & 26.500\\
Category vector & 7 & 0 & 0 & 0.0000 & 4.250 & 1.208 & 2.708\\
Mixed radix & 3 & 0 & 0 & 0.0000 & 7.333 & 1.708 & 5.459\\
Catalogue statistic & 2 & 0 & 0 & 0.0000 & 14.625 & 2.875 & 4.834\\
Universal bit code & 4 & 4 & 1 & 0.1322 & 54.167 & 39.875 & 143.959\\
Raw cache, 0 B & 0 & 4 & 4 & 1.2382 & 10.042 & 10.500 & 35.375\\
Raw cache, 1 B & 1 & 4 & 3 & 0.8083 & 7.375 & 9.000 & 36.167\\
Raw cache, 2 B & 2 & 4 & 2 & 0.2827 & 6.875 & 3.958 & 32.541\\
Raw cache, 3 B & 3 & 4 & 1 & 0.1383 & 6.708 & 3.958 & 21.042\\
Raw cache, 4 B & 4 & 0 & 0 & 0.0000 & 6.083 & 3.917 & 9.417\\
\bottomrule\end{tabular}

\caption{Native-source resource costs. B denotes bytes, $R_{\max}$ is the cap
on byte reads per query, and p95 is the query-latency 95th percentile.}
\label{tab:agent-native}
\end{table}

The missing-archive and stale-archive conditions distinguish abstention
from changes to the original source decisions.
Appendix~\ref{app:agent} reports these outcomes, the history-source
resource comparison, query-population shift, and shared extraction costs.
Intervals resample complete dialogues and retain all their entity queries
together. Decoder timings include compatibility evaluation.
The upstream measurements separately charge text processing and updating.

\subsection{Computation and physical resources}
Table~\ref{tab:timing} reports median writing and query latency for the bit
implementation. Generic query time includes candidate reconstruction
and predicate evaluation. The largest one-read record occupies the
same rounded payload size as full-record storage. Logical bit savings
therefore do not necessarily reduce the physical payload.

We separately measured Python allocation peaks around each writer and
reader call. These measurements excluded pre-existing inputs, archives,
interpreter memory, native allocations, and operating-system resident
memory. Table~\ref{tab:protocol} records the allocation sample size and
shared dialogue resources. The word implementation charges complete
word responses and is checked separately from the bit timings.

\begin{table}[t]
\centering\small\begin{tabular}{rrrrrrr}\toprule
$n$ & $R$ & Bits & Bytes & Write ($\mu$s) & Query ($\mu$s) & Query p95 ($\mu$s)\\\midrule
10 & 1 & 8 & 1 & 7.5 & 14.2 & 14.8\\
10 & 4 & 5 & 1 & 6.6 & 14.5 & 15.0\\
10 & 10 & 0 & 0 & 3.2 & 5.6 & 6.1\\
64 & 1 & 62 & 8 & 29.2 & 21.3 & 22.2\\
64 & 4 & 59 & 8 & 28.0 & 21.3 & 22.5\\
64 & 16 & 47 & 6 & 23.5 & 22.5 & 23.2\\
256 & 1 & 253 & 32 & 114.2 & 364.5 & 375.7\\
256 & 4 & 250 & 32 & 111.0 & 361.0 & 368.1\\
256 & 16 & 238 & 30 & 109.6 & 365.7 & 370.2\\
1024 & 1 & 1021 & 128 & 484.8 & 510.4 & 517.9\\
1024 & 4 & 1018 & 128 & 464.4 & 496.9 & 511.1\\
1024 & 16 & 1006 & 126 & 488.8 & 526.1 & 534.2\\
\bottomrule\end{tabular}

\caption{Median bit-codec latency, query 95th percentile (p95), and retained state.}
\label{tab:timing}
\end{table}

\section{Limitations}
The theorems quantify over full-domain records and all Boolean queries,
including long truth tables. Restricted sources or query families can
admit smaller sufficient statistics. The results do not establish
comparable hardness for natural-language questions.
The archive must contain unchanged raw coordinates or disjoint words,
and all record-dependent metadata is charged. Writing and decoding can
require substantial transient storage and computation. Query purity is
a caller contract. Finite computational checks do not replace the proofs.

The codecs do not detect staleness or repair extraction errors.
The matched evaluation uses previously trained linear text models on
a source split already exposed in Nous research. The full-history
source, catalogue study, and constant controls were added after the
initial native-source outcomes.
The task validates selected categorical constraints and does not measure
complete dialogue success. The cached relation control contains
constructed questions whose imbalance can raise accuracy.
Decoder timings and Python call allocations exclude the observation stage.
We measure shared extraction costs and upstream resident memory separately.

\section{Conclusion}
Pre-query retention requires a writer to partition records into cells
that a bounded reader can resolve. We characterize one-bit-read cells,
construct full-domain encoders, and derive universal retention bounds
for bit and word retrieval. Word width determines whether nonlinear
retention provides any saving beyond raw caching. An explicit regime
makes raw-word caching exactly optimal, including a range of positive
average-error allowances. The matched evaluation favors a categorical
sufficient statistic in storage and decoding latency. The catalogue index
trades a smaller private state for shared storage and selective query coverage.
Together, the results connect retention requirements to the retrieval
interface, record support, and future query family.

\begin{acks}
Artificial intelligence tools assisted with implementation and manuscript preparation.
The authors received no funding for this work.
The authors declare no competing interests.
\end{acks}
\printbibliography

@article{permuter2011,
 author={Permuter, Haim H. and Weissman, Tsachy},
 title={Source Coding with a Side Information ``Vending Machine''},
 journal={IEEE Transactions on Information Theory},volume={57},number={7},pages={4530--4544},year={2011},
 doi={10.1109/TIT.2011.2145510},url={https://arxiv.org/abs/0904.2311}}

@misc{makhdoumi2013,
 author={Makhdoumi, Ali and Huang, Shao-Lun and M{\'e}dard, Muriel and Polyanskiy, Yury},
 title={On Locally Decodable Source Coding},year={2013},eprint={1308.5239},archivePrefix={arXiv},
 url={https://arxiv.org/abs/1308.5239v2}}

@inproceedings{ramamoorthy2020,
 author={Natarajan Ramamoorthy, Sivaramakrishnan and Rashtchian, Cyrus},
 title={Equivalence of Systematic Linear Data Structures and Matrix Rigidity},
 booktitle={11th Innovations in Theoretical Computer Science Conference},
 series={LIPIcs},volume={151},pages={35:1--35:20},year={2020},
 doi={10.4230/LIPIcs.ITCS.2020.35},url={https://drops.dagstuhl.de/entities/document/10.4230/LIPIcs.ITCS.2020.35}}

@misc{wen2024,
 author={Wen, Kaiyue and Dang, Xingyu and Lyu, Kaifeng},
 title={{RNNs} are not Transformers (Yet): The Key Bottleneck on In-context Retrieval},
 year={2024},eprint={2402.18510},archivePrefix={arXiv},url={https://arxiv.org/abs/2402.18510v4}}

@misc{zou2026,
 author={Zou, Mingxi and Guo, Zhihan and Liang, Langzhang and Wang, Zhuo and Wang, Qifan and Wen, Qingsong and King, Irwin and Qu, Lizhen and Xu, Zenglin},
 title={Remember the Decision, Not the Description: A Rate-Distortion Framework for Agent Memory},
 year={2026},eprint={2605.10870},archivePrefix={arXiv},url={https://arxiv.org/abs/2605.10870}}

@misc{singh2026nous,
 author={Singh, Pranav},title={Nous: A Predictive World Model for Long-Term Agent Memory},
 year={2026},eprint={2606.22030},archivePrefix={arXiv},url={https://arxiv.org/abs/2606.22030}}

@misc{singh2026certification,
 author={Singh, Pranav},title={Nous: Learning and Certifying Memory Decisions Before Source Calibration},
 year={2026},eprint={2610.00094},archivePrefix={arXiv},note={Version 2},url={https://arxiv.org/abs/2610.00094v2}}

@misc{nousrelease,
 author={Singh, Pranav},title={Nous: Public Research-Only Reproducibility Snapshot},year={2026},
 note={Release jmlr-submission-2026-10-04},
 url={https://github.com/Pranavsingh431/nous-state/releases/tag/jmlr-submission-2026-10-04}}

@inproceedings{multiwoz24,
 author={Ye, Fanghua and Manotumruksa, Jarana and Yilmaz, Emine},
 title={{MultiWOZ 2.4}: A Multi-Domain Task-Oriented Dialogue Dataset with Essential Annotation Corrections to Improve State Tracking Evaluation},
 booktitle={Proceedings of the 23rd Annual Meeting of the Special Interest Group on Discourse and Dialogue},
 publisher={Association for Computational Linguistics},year={2022},pages={351--360},
 doi={10.18653/v1/2022.sigdial-1.34},url={https://aclanthology.org/2022.sigdial-1.34/}}

@inproceedings{yildiz2012,
 author={Y{\i}ld{\i}z, Olcay Taner},
 title={On the {VC}-dimension of Univariate Decision Trees},
 booktitle={Proceedings of the 1st International Conference on Pattern Recognition Applications and Methods},
 volume={1},pages={205--210},year={2012},publisher={SciTePress},
 doi={10.5220/0003777202050210},
 url={https://www.scitepress.org/PublishedPapers/2012/37772/}}

@article{yildiz2015,
 author={Y{\i}ld{\i}z, Olcay Taner},title={{VC}-Dimension of Univariate Decision Trees},
 journal={IEEE Transactions on Neural Networks and Learning Systems},
 volume={26},number={2},pages={378--387},year={2015},
 doi={10.1109/TNNLS.2014.2385837}}

@misc{leboeuf2022,
 author={Leboeuf, Jean-Samuel and LeBlanc, Fr{\'e}d{\'e}ric and Marchand, Mario},
 title={Generalization Properties of Decision Trees on Real-valued and Categorical Features},
 year={2022},eprint={2210.10781},archivePrefix={arXiv},
 url={https://arxiv.org/abs/2210.10781}}

@techreport{gal2003,
 author={G{\'a}l, Anna and Miltersen, Peter Bro},
 title={The Cell Probe Complexity of Succinct Data Structures},
 institution={BRICS, University of Aarhus},number={RS-03-44},year={2003},
 url={https://www.brics.dk/RS/03/44/BRICS-RS-03-44.pdf}}

@article{nisan1999,
 author={Nisan, Noam and Rudich, Steven and Saks, Michael},
 title={Products and Help Bits in Decision Trees},
 journal={SIAM Journal on Computing},volume={28},number={3},
 pages={1035--1050},year={1999},doi={10.1137/S0097539795282444}}

@article{viola2012,
 author={Viola, Emanuele},
 title={Bit-Probe Lower Bounds for Succinct Data Structures},
 journal={SIAM Journal on Computing},volume={41},number={6},
 pages={1593--1604},year={2012},doi={10.1137/090766619},
 url={https://epubs.siam.org/doi/10.1137/090766619}}

@inproceedings{patrascu2010,
 author={P{\u{a}}tra{\c{s}}cu, Mihai and Viola, Emanuele},
 title={Cell-Probe Lower Bounds for Succinct Partial Sums},
 booktitle={Proceedings of the Twenty-First Annual ACM-SIAM Symposium on Discrete Algorithms},
 publisher={Society for Industrial and Applied Mathematics},
 pages={117--122},year={2010},doi={10.1137/1.9781611973075.11},
 url={https://people.csail.mit.edu/mip/papers/ranklb/paper.pdf}}

@article{bilgic2011,
 author={Bilgic, Mustafa and Getoor, Lise},
 title={Value of Information Lattice: Exploiting Probabilistic Independence for Effective Feature Subset Acquisition},
 journal={Journal of Artificial Intelligence Research},volume={41},pages={69--95},
 year={2011},doi={10.1613/jair.3200},url={https://arxiv.org/abs/1401.3881}}

@article{subramanian2022,
 author={Subramanian, Jayakumar and Sinha, Amit and Seraj, Raihan and Mahajan, Aditya},
 title={Approximate Information State for Approximate Planning and Reinforcement Learning in Partially Observed Systems},
 journal={Journal of Machine Learning Research},volume={23},number={12},
 pages={1--83},year={2022},url={https://jmlr.org/papers/v23/20-1165.html}}
\appendix
\section{Proof of the structural characterization}\label{app:structure}
We first justify the decoding characterization used throughout.
On a nonempty cell $S$, the reader's state and query are fixed before
its first response. Its first address is therefore fixed on the entire
cell. Each Boolean function of the response is constant, identity, or
complement. Thus its output restriction lies in $\mathcal D(S)$.
Conversely every member of this set is obtainable with zero or one read.

Fix $x^0\in S$, let $V=\operatorname{span}(S+x^0)$, and write
$\ell_i(v)=v_i$ for the coordinate restrictions to $V$.
Suppose all pairwise XORs are available. Then, for distinct $i,j$,
$c_i+c_j$ is constant or a coordinate plus a constant on $S$.
Subtracting its value at $x^0$ implies that $\ell_i+\ell_j$ is zero
or a listed $\ell_a$ on the spanning set $S+x^0$, hence on $V$.
The same-index sum is zero. Therefore
$E=\{0,\ell_1,\ldots,\ell_n\}$ is additively closed.
The coordinate forms span $V^*$: any linear functional on $V$ extends
to $\F^n$ and is a sum of ambient coordinates. Hence $E=V^*$.
Conversely, if $E=V^*$, each $\ell_i+\ell_j$ is zero or a coordinate
restriction, and adding the appropriate value at $x^0$ recovers
$c_i+c_j$ on $S$. The same reasoning holds on $x^0+V$.
Finally, $2^{\dim V}=|E|\le n+1$, proving part one.

Now assume all pairwise XOR and AND functions are in $\mathcal D(S)$.
XOR feasibility makes this set additively closed: constants and
complements are included, and any sum of two coordinate literals
reduces to a pair XOR or a constant.
For distinct coordinates,
\begin{equation}
 (c_i+a)(c_j+b)=c_ic_j+bc_i+ac_j+ab,\qquad a,b\in\F,
\end{equation}
belongs to $\mathcal D(S)$ by additive closure and pairwise AND
availability. A product involving a constant or two literals of the same
coordinate is a constant or a unary function and is already available.
Consequently $\mathcal D(S)$ is closed under products and complements,
and therefore under all Boolean operations.
For each row $u\in S$, the product of the literals matching $u$
over all coordinates is the singleton indicator of $u$.
The coordinates distinguish distinct rows, so this product vanishes
at every other row. Sums of singleton indicators give every Boolean
function on $S$. Thus XOR-plus-AND availability implies universality.
Universality implies every pair gate, and the latter implies XOR-plus-AND.

To characterize universality geometrically, assume XOR feasibility and
put $d=\dim V$. The set $\mathcal D(S)$ then consists of
restrictions of all affine functions on $x^0+V$. There are $2^{d+1}$
such functions, with distinct restrictions: an affine function vanishing
on $S$ vanishes at $x^0$ and on the linear span $V$.
Hence $\mathcal D(S)$ contains all $2^{|S|}$ Boolean functions on $S$
if and only if $d+1=|S|$, which is affine independence.
These implications prove all equivalences in Theorem~\ref{thm:structure}.
In particular, the normalized columns on the other $|S|-1$ rows must
include every nonzero binary vector, so
$n\ge 2^{|S|-1}-1$ and $|S|\le s_n$.
The resulting bound is the binary-stump capacity attributed above.

\section{Proof of the exact optima and endpoints}\label{app:endpoints}
For XOR, the structural result gives $|S|\le2^r$ for
$r=\lfloor\log_2(n+1)\rfloor$. A $B$-bit writer partitions $2^n$
records into at most $2^B$ cells. Thus $2^n\le2^B2^r$ and $B\ge n-r$.
The anchor encoder in Section~\ref{sec:xor-code} is a total map on the
cube: every record supplies its anchor bits and every nonanchor supplies
one offset. Conversely, any anchor vector $z$ and any offset payload
recover exactly one record. Every payload therefore defines $2^r$
records and every payload is used. The stated decoder algebra proves
correctness with at most one read. The encoder attains the lower bound.

For the universal workload, $|S|\le s_n$ gives
$B\ge\lceil n-\log_2 s_n\rceil=n-\lfloor\log_2 s_n\rfloor$.
Set $k=\lfloor\log_2 s_n\rfloor$ and $t=2^k$.
There are $2^{t-1}-1$ nonzero Boolean functions on $\bits^k$
vanishing at zero. Since $t\le s_n$,
\begin{equation}
 2^{t-1}-1\le2^{\lfloor\log_2(n+1)\rfloor}-1\le n .
\end{equation}
Their identity features provide distinct anchors. For every anchor
vector and every nonanchor offset vector the formula
$x_i=a_i+\phi_i(z)$ defines one record, whose anchor values are
exactly $z$ and whose encoder returns exactly those offsets.
Thus this is a partition into $2^{n-k}$ cells of size $2^k$.
For any Boolean query, its normalized candidate truth table occurs in
the public feature list unless it is zero. The decoder in
Section~\ref{sec:bool-code} is therefore defined and correct on every
cell. The construction proves Equation~\eqref{eq:universal}, for pair gates and
arbitrary queries alike.

For affine one-read writers, every nonempty cell is an affine space
$x^0+L$. Universal feasibility forces it to be affinely independent.
An affine space of dimension $d$ has $2^d$ points and affine dimension
$d$, so affine independence requires $2^d=d+1$, hence $d\le1$.
A writer with $B<n-1$ affine output bits has a kernel of dimension
at least two and violates this condition. Retaining $n-1$ raw
coordinates and reading the remaining one attains the bound.

At zero reads, two records are indistinguishable for all pair XORs if
and only if they differ by the all-ones vector or are equal.
Indeed equality of $x_1+x_j$ for every $j>1$ fixes every difference
relative to coordinate one. There are $2^{n-1}$ equivalence classes.
Retaining those $n-1$ differences attains the bound. Boolean pair
queries include all coordinate projections for $n\ge2$, so a no-read
writer must be injective and requires $n$ bits. Two reads directly
obtain the two queried coordinates for either pair workload, yielding
zero retained bits. For arbitrary Boolean queries, zero reads likewise
require $n$ bits and $n$ reads suffice with zero bits.

If $n=1$, the pair workload is empty and needs zero bits. The arbitrary
one-bit workload can be answered by one raw read. The binary codec classes
restrict their dimensions to $n\ge2$. Empty records can be treated
separately with zero retained bits. These edge conventions are not used
to extrapolate the displayed pair formulas outside their stated domain.

For zero-error randomization independent of the record and query, fix
one random tape for the writer and a tape for each possible query.
There are finitely many records and Boolean functions on the finite
cube. Each correctness event has probability one, so their finite
intersection has probability one. A tape in that intersection supplies
a deterministic scheme with the same fixed state and worst-case read
constraints. The deterministic lower bounds apply. This argument does
not apply to positive-error guarantees or expected read budgets.

\section{Proof of the multiple-read bounds}\label{app:multi}
Condition on one retained state. A deterministic reader with at most
$R$ coordinate reads is a binary decision tree of depth at most $R$.
The tree can be padded to a full depth-$R$ tree by repeating arbitrary
coordinate reads below terminated leaves and preserving their outputs.
There are $2^R-1$ internal nodes, each with at most $n$ address choices,
and $2^R$ leaves, each with two output choices. Thus there are at most
\begin{equation}
 T_{n,R}=n^{2^R-1}2^{2^R}=2^{C_{n,R}}
\end{equation}
tree descriptions. The count overcounts functions and unreachable branches,
which is safe for a converse.
Every Boolean labeling on a cell $S$ extends to a query on the cube.
Universal exact answerability therefore requires
$2^{|S|}\le T_{n,R}$, or
$|S|\le A=\lfloor C_{n,R}\rfloor$.
Since $2^n\le2^B A$, the lower bound in Equation~\eqref{eq:multi}
follows, clamped at zero. The sharper exact one-read bound takes
precedence when $R=1$.

For the upper bound put $N=n-R+1$ and
$k=\lfloor\log_2(1+\lfloor\log_2(N+1)\rfloor)\rfloor$.
Reserve $R-1$ raw identity coordinates as selector anchors, and use
the universal feature list on the remaining $N$ coordinates with $k$
tail anchors. Retain all $N-k=n-R+1-k$ nonanchor offsets.
Every selector vector, tail-anchor vector, and offset vector specifies
exactly one full record. The bijection proves that the writer partitions the
whole cube and that no input-dependent index is hidden.
The reader obtains the selector vector with $R-1$ charged reads.
For each of the $2^k$ tail vectors, it reconstructs the corresponding
hypothetical full record, using those selectors and the offsets.
A single final feature read recovers the query value as before.
All earlier selector responses are charged, and the last address
may depend on them. At $R=n$, $N=1$ and $k=1$, so the retained
state is empty and the construction uses at most $n$ reads.

For the uniform constants, let $L=\log_2 n\ge1$.
We have $C_{n,R}\le2^R(1+L)\le2^{R+1}L$.
The converse hence gives $B\ge n-R-\log_2 L-1$.
If $R\le n/2$, then $n-R+2\ge n/2+2\ge\sqrt n$, and
$1+\lfloor\log_2(n-R+2)\rfloor>\tfrac12\log_2 n$.
Taking a logarithm and flooring yields
$k>\log_2 L-2$. The upper construction gives
$B=n-R+1-k<n-R-\log_2 L+3$, proving the stated safe bound.

Finally, suppose an affine writer has kernel $L$ of dimension $d>R$,
and consider a nonempty cell $x^0+L$.
Use the query $q(x)=\mathbf1\{x=x^0\}$.
The reader path taken by $x^0$ imposes at most $R$ affine coordinate
equations on that cell. Its solution set has dimension at least
$d-R\ge1$, so some $x^1\ne x^0$ follows exactly the same path.
The same leaf cannot be correct for both point-query labels.
Thus $d\le R$. Since $\operatorname{rank} W\le B$, this gives
$B\ge n-R$. Retaining $n-R$ raw bits and reading all remaining $R$
bits attains equality for arbitrary queries.
The point-query argument does not give that lower bound for pair-only
workloads.

\section{Proof of the average-error converse}\label{app:approx}
Replace every abstention by the binary answer zero. The replacement cannot increase
the error probability, since abstention was counted as an error. We can therefore prove a lower bound for binary-output readers.
Let $H$ denote Shannon entropy and $I$ mutual information.
Let $N=2^n$. Choose a uniformly random Boolean query by assigning
independent fair labels $Y_x$ to all $x\in\bits^n$.
Use a fixed canonical truth-table description for each query.
Let $U$ contain all random tapes, independent of this query and of
the record. Granting all these tapes to the reader only makes its task
easier. For each fixed tape, the writer is a deterministic map with at
most $K=2^B$ state labels.

For every state label, record the depth-at-most-$R$ tree selected by the
reader for this query and tape, padded as above. Put all $K$ tree
descriptions into a vector $T$. Unused state labels receive a fixed
dummy tree. The number of possibilities for $T$, conditioned on $U$,
is at most $(T_{n,R})^K$. Hence
\begin{equation}
 I(Y;T\mid U)\le H(T\mid U)\le K C_{n,R}.
\end{equation}
The tree may depend on the entire query description, including labels
outside its cell.

Given $U,T$, reconstruct all $N$ labels by running, at each record $x$,
the tree indexed by $W_U(x)$. These are hypothetical evaluations in
the proof. Write this reconstruction as $\widehat Y$ and set
$E_x=Y_x+\widehat Y_x$. Because $\widehat Y$ is a deterministic function
of $U,T$, conditional translation gives
\begin{equation}
 H(Y\mid T,U)=H(E\mid T,U)\le H(E)
 \le\sum_x h_2(p_x)\le N h_2(e),
\end{equation}
where $p_x=\Pr\{E_x=1\}$ and
$e=N^{-1}\sum_xp_x$. The last inequality is concavity of binary entropy.
Independence and fairness give $H(Y\mid U)=N$, and therefore
\begin{equation}
 N(1-h_2(e))\le I(Y;T\mid U)\le 2^B C_{n,R}.
\end{equation}
The assumed error bound holds for every fixed query, so averaging over
the random query yields $e\le\epsilon<1/2$.
Binary entropy is increasing on $[0,1/2]$, proving
$2^B C_{n,R}\ge2^n(1-h_2(\epsilon))$.
Taking logarithms and using integer $B\ge0$ gives
Theorem~\ref{thm:approx}. The exact construction supplies its upper
bound. Combining it with $C_{n,R}\le2^{R+1}\log_2 n$ yields the
uniform $O_\epsilon(1)$ statement. No fresh-sample or independence
assumption about adaptively retrieved archive values has been used.

\section{Reproduction and resource details}\label{app:reproduction}
The reproduction driver runs the exhaustive checks,
controlled simulations, cached Nous replay, derived relation queries,
result audit, and figure generation. The reference environment is
Python 3.12.14, NumPy 2.5.4, and Matplotlib 3.11.2 on macOS with an
Apple M1 CPU and 8 GB RAM. No GPU or paid inference is used.
CPU cache sizes were not available to the sandbox. Timings depend on
local load and are not expected to be bitwise reproducible.
Seeds and all per-cell sample counts are recorded in result JSON files.

The controlled source uses Python's Mersenne Twister with seed 15102026.
Each world is a uniform $n$-bit integer. Pair indices are sampled
uniformly without replacement. Masked queries draw a uniform $n$-bit
mask and replace a zero draw with one (so the final mask is not uniform
over nonzero masks). Thresholds
are the ceiling of half the mask weight. Corruption independently
flips each archive bit at the stated probability, after writing.
Ground truth remains the pre-corruption query value.
The removed condition makes the read callback unavailable. A decoder
returns abstention when it needs a read. No archive-free cache of
previous queries is provided.

The matched raw cache stores the first $B$ raw bits and enumerates all
$2^{n-B}$ completions. A one-step selective decision tree chooses the
coordinate maximizing the number of candidates in answer-homogeneous
branches, with the lowest-coordinate tie break. Branches with mixed
labels abstain. If the query is constant over all candidates, no read
is taken. This policy attains optimal selective coverage for that cell under a
uniform source, at one probe. The statement fixes the writer.

Dialogue queries use seed 15102027. For each dialogue, distinct feature pairs and gates are sampled
using the counts in Table~\ref{tab:protocol}. Identical queries
are applied to every source method. Gates use truth-table
index $2x_i+x_j$. Dialogue relation intervals use whole-dialogue bootstrap replicates
with NumPy seed 15102028. The slot replay
uses seed 14102026. Methods are not selected by test performance.
The data and module hashes accompany the reproducibility package.

Each per-record codec payload contains exactly $\lceil B/8\rceil$
bytes with zero padding, no header or embedded dimension.
The dimension and feature order must be supplied by the public schema.
A real container adds framing, identifiers, and storage overhead.
The mathematical archive uses $n$ bits. The callback implementation
uses $n$ byte-valued Boolean facts before Python object overhead.
Its measurement address log is not an input to later queries.
Shared source code, fixed schemas, original model bytes, archive
bytes, batch-writer workspace, and query workspace are not included
in the retained bit count. No history-specific model fit or embedding
index is created by these experiments.

The word-codec verification uses seed 16102026 for sampled threshold
queries. Its exhaustive checks cover every truth table on the stated
small cubes, followed by parity and point queries on larger cubes.
The accompanying results specify each dimension and read budget.
The bound checker uses exact integer arithmetic when the word count
is a power of two. These checks preserve the original bit-experiment
results and timings.

\section{Proofs for word retrieval}\label{app:words}
For a fixed memory cell, pad a reader to a full depth-$R$ tree with
$Q$ branches per internal node. The padded tree has $Q^R$ leaves and
$(Q^R-1)/(Q-1)$ internal nodes. Each internal node chooses among $M$
word addresses. Each leaf chooses a binary output. Thus the number
of descriptions is at most
\begin{equation}
 T=M^{(Q^R-1)/(Q-1)}2^{Q^R}=2^{C^{(w)}_{M,R}}.
\end{equation}
The count includes unreachable branches and repeated addresses.
Every labeling of a cell extends to a Boolean query on the full record.
Consequently each cell has at most $\lfloor C^{(w)}_{M,R}\rfloor$ records.
The writer partitions $2^N$ records into at most $2^B$ cells, which gives
Equation~\eqref{eq:word-lower}.

A raw-word cache retains $N-wR$ bits and reconstructs the record with
$R$ reads. For the alternative, reserve $R-1$ complete words as selector
anchors. The remaining $s=w(M-R+1)$ bits receive the bit encoder of
Section~\ref{sec:bool-code}, with $k=k^{(w)}_{M,R}$ anchors.
The writer retains only the $s-k$ nonanchor offsets.
Every selector vector, tail-anchor vector, and offset vector defines
exactly one record. The construction therefore partitions the full cube.
After reading the selectors, the reader evaluates the query on each
tail candidate. The selected bit is obtained from its containing word.
The other bits in that response are ignored. The total is at most $R$
word reads, giving Equation~\eqref{eq:word-upper} after comparison with
the raw-word cache. When $s=1$, the tail is read directly.

If $\log_2 M\le Q-1$, then
\begin{equation}
 C^{(w)}_{M,R}\le Q^R+(Q^R-1)=2Q^R-1<2^{wR+1}.
\end{equation}
The count implies that $\lfloor\log_2\lfloor C^{(w)}_{M,R}\rfloor\rfloor\le wR$.
The converse and raw-word cache prove Equation~\eqref{eq:word-exact}.
For $R=0$, coordinate queries force an injective writer, so $B=N$.
Zero-error randomness can be fixed by the finite-intersection argument
in Appendix~\ref{app:endpoints}.

To prove the uniform gap, let $L=\log_2 M\ge1$,
$a=\lfloor\log_2\lfloor L\rfloor\rfloor$, and
$d=\max\{0,a-w\}$. For $R\le M/2$,
$s+1=w(M-R+1)+1\ge M/2+2$. Therefore
$1+\lfloor\log_2(s+1)\rfloor\ge\lfloor L\rfloor$ and $k\ge a$.
The constructive retained length is at most $N-wR-d$.
Set $A=\log_2(1+L/(2^w-1))$. Since
$C^{(w)}_{M,R}\le2^{wR}(1+L/(2^w-1))$, the integer lower bound is at
least $N-wR-\lfloor A\rfloor$.

The definition of $a$ gives $L<2^{a+1}$.
If $a\le w$, then $d=0$ and
$A<\log_2(1+2^{w+1}/(2^w-1))\le\log_2 5<3$.
If $a>w$, then
\begin{equation}
 A-d=\log_2\!\left(2^{-d}+\frac{L}{(2^w-1)2^d}\right)
 <\log_2\!\left(1+\frac{2^{w+1}}{2^w-1}\right)\le\log_2 5<3.
\end{equation}
Hence $\lfloor A\rfloor-d\le2$ in both cases.
The difference between the constructive and converse bounds is at most
two retained bits. Clamping the lower bound at zero cannot increase
that difference.

For Proposition~\ref{prop:word-error}, replace the binary-tree count in
Appendix~\ref{app:approx} by $T=2^{C^{(w)}_{M,R}}$ and use $2^N$ random
query labels. The same conditional entropy proof gives
\begin{equation}
 2^N(1-h_2(\epsilon))\le2^B C^{(w)}_{M,R}.
\end{equation}
The proof permits shared or private independent random tapes and counts
abstention as error. Taking logarithms gives Equation~\eqref{eq:word-error}.
Condition~\eqref{eq:word-error-regime} makes its unrounded lower bound
strictly greater than $N-wR-1$. Integer retention therefore requires
$B\ge N-wR$, and the exact cache attains equality.

Finally suppose $h=h_2(\epsilon)<1/2$ and
$\log_2 M\le(Q-1)(1-2h)$.
If $M>1$, the left side of Equation~\eqref{eq:word-error-regime} is
strictly smaller than $1+(1-2h)=2(1-h)$ because $R\ge1$.
If $M=1$, that left side equals one, which is also strictly smaller
than $2(1-h)$. These cases prove the sufficient condition.

\section{Matched evaluation details}\label{app:agent}
The study reuses the original pretrained text models without fitting or selection.
The local source uses term-frequency and inverse-document-frequency
features from the user turn and preceding system turn.
The history source uses the original combined history and local features.
The model hashes and original training lock accompany the package.
We checked that official test dialogue identifiers are disjoint from
the original training and development identifiers.

The writer receives only the source categorical probabilities.
It receives neither the selected catalogue entity nor benchmark metadata.
The source models never receive future system text or state annotations.
A dialogue is excluded only when its final target contains a value
outside the fixed ontology. Table~\ref{tab:agent-protocol} records the
included population. A query checks only the slots tracked in this study.
An unspecified or indifferent constraint accepts every catalogue value.
Other constraints require an exact canonical match.

The native-source method plan was fixed before computing the new offer
outcomes. The history-source sensitivity analysis and both constant
controls were added after the native-source outcomes were inspected.
The catalogue baseline and attribute-grid control were added after
both source groups had been evaluated.
The primary population remained unchanged. We report every source and control.
No source model was refitted or selected by these outcomes.
The public schema and query predicates determine the catalogue code.
Protocol versions and hashes accompany the results.

Every writer returns a fixed byte payload with public dimensions.
The mixed-radix encoding uses the ontology ordering.
The universal encoding normalizes query labels over its candidate records.
The raw caches retain complete words in a public suffix.
A raw reader maintains the set of values consistent with its observed
indicator bits. It returns an answer when every compatible value agrees.
Otherwise it reads the lowest-index uncached word relevant to the query.
A contradictory categorical observation causes abstention.
Each query begins without a transcript from previous queries.

\begin{table}[t]
\centering\scriptsize\begin{tabular}{lrrrrrrr}\toprule
Method & State (B) & Archive (B) & $R_{\max}$ & Returned (B/query) & Write ($\mu$s) & Query ($\mu$s) & p95 ($\mu$s)\\\midrule
Posterior summary & 312 & 0 & 0 & 0.0000 & 15.791 & 5.500 & 11.291\\
Category vector & 7 & 0 & 0 & 0.0000 & 2.000 & 1.000 & 1.625\\
Mixed radix & 3 & 0 & 0 & 0.0000 & 3.583 & 1.208 & 3.375\\
Catalogue statistic & 2 & 0 & 0 & 0.0000 & 11.875 & 2.167 & 4.000\\
Universal bit code & 4 & 4 & 1 & 0.1202 & 38.667 & 34.458 & 55.916\\
Raw cache, 0 B & 0 & 4 & 4 & 1.3297 & 4.416 & 7.833 & 20.209\\
Raw cache, 1 B & 1 & 4 & 3 & 0.6859 & 2.708 & 7.542 & 21.292\\
Raw cache, 2 B & 2 & 4 & 2 & 0.2518 & 2.291 & 3.250 & 20.792\\
Raw cache, 3 B & 3 & 4 & 1 & 0.1126 & 2.167 & 3.250 & 13.167\\
Raw cache, 4 B & 4 & 0 & 0 & 0.0000 & 2.166 & 3.000 & 5.875\\
\bottomrule\end{tabular}

\caption{History-source resource costs. B denotes bytes, $R_{\max}$ is the cap
on byte reads per query, and p95 is the query-latency 95th percentile.}
\label{tab:agent-history}
\end{table}

Table~\ref{tab:agent-history} gives the second source's resource measurements.
Table~\ref{tab:agent-failure} reports missing-archive coverage and disagreement
with the original source under corruption. We independently flipped
archived bits after writing and kept the retained state unchanged.
Every method encountered the same corrupted archive within a dialogue.
The state-only representations preserve their original answers in both
conditions because they require no archive response.

\begin{table}[t]
\centering\small\begin{tabular}{llrrr}\toprule
Source & Method & Removed coverage & Stale coverage & Stale disagreement\\\midrule
Native & Mixed radix & 1.0000 & 1.0000 & 0.0000\\
Native & Catalogue statistic & 1.0000 & 1.0000 & 0.0000\\
Native & Universal bit code & 0.8678 & 1.0000 & 0.0009\\
Native & Raw cache, 1 B & 0.3274 & 0.9527 & 0.0084\\
Native & Raw cache, 2 B & 0.8530 & 0.9848 & 0.0027\\
Native & Raw cache, 3 B & 0.8617 & 0.9937 & 0.0017\\
History & Mixed radix & 1.0000 & 1.0000 & 0.0000\\
History & Catalogue statistic & 1.0000 & 1.0000 & 0.0000\\
History & Universal bit code & 0.8798 & 1.0000 & 0.0021\\
History & Raw cache, 1 B & 0.4189 & 0.9724 & 0.0101\\
History & Raw cache, 2 B & 0.8530 & 0.9909 & 0.0037\\
History & Raw cache, 3 B & 0.8874 & 0.9959 & 0.0013\\
\bottomrule\end{tabular}

\caption{Archive-failure coverage and disagreement with source decisions.
Stale disagreement is conditional on answering.}
\label{tab:agent-failure}
\end{table}

The catalogue compiler enumerates each domain's categorical states and
groups states with identical answer patterns. Its reader decodes the
retained pattern index and returns the indexed entity answer.
Every combination of domain patterns is realizable on the product support.
The number of distinct complete answer vectors is therefore the product
of the domain counts in Table~\ref{tab:agent-protocol}.
Two states with different vectors must have different zero-read payloads.
This pigeonhole argument proves Equation~\eqref{eq:catalogue}.

For query-family shift, we construct an attribute grid from the public
ontology. The grid includes profiles absent from the observed catalogue.
These are controlled compositional queries on the same dialogue-derived
states. The catalogue reader evaluates each new predicate on
all states in its retained equivalence class. It answers only when every
state agrees. The mixed-radix reader answers the same queries from its
complete categorical state. Table~\ref{tab:catalogue-shift} reports coverage
and conditional error on profiles absent from the catalogue.

\begin{table}[t]
\centering\small\begin{tabular}{llrrr}\toprule
Source & Method & Queries & Coverage & Conditional error\\\midrule
Native & Catalogue statistic & 12,900 & 0.9606 & 0.7334\\
Native & Mixed radix & 12,900 & 1.0000 & 0.7244\\
History & Catalogue statistic & 12,900 & 0.8663 & 0.1774\\
History & Mixed radix & 12,900 & 1.0000 & 0.1952\\
\bottomrule\end{tabular}

\caption{Selective answering on unseen attribute profiles.}
\label{tab:catalogue-shift}
\end{table}

Table~\ref{tab:agent-shift} compares the primary population with queries
from every domain. The wider population contains domains without a
tracked constraint. Its accuracy therefore describes a different workload.
We retain the constrained population as the primary comparison.

\begin{table}[t]
\centering\small\begin{tabular}{llrrr}\toprule
Source & Query population & Queries & Accuracy & Balanced accuracy\\\midrule
Native update & Active constraints & 67,346 & 0.4725 & 0.6170\\
Native update & All domains & 221,778 & 0.8318 & 0.6322\\
History classifier & Active constraints & 67,346 & 0.8070 & 0.8461\\
History classifier & All domains & 221,778 & 0.9331 & 0.8679\\
\bottomrule\end{tabular}

\caption{Accuracy under a change of query population.}
\label{tab:agent-shift}
\end{table}

Table~\ref{tab:agent-upstream} records the shared observation costs.
The native source predicts at each user turn and updates its beliefs.
The history source predicts once per final dialogue prefix.
Both source groups load the same pretrained model bundle.
Catalogue descriptors are public query inputs prepared before timing.
The decoder timer covers retained-state decoding and compatibility
checking, while the model timer covers the shared extraction stage.

\begin{table}[t]
\centering\small\begin{tabular}{lrrrrr}\toprule
Source & Predictions & Text (B) & Extraction (s) & Update (s) & Load (s)\\\midrule
Native update & 51,604 & 1,094,742 & 0.639 & 1.134 & 3.014\\
History classifier & 7,000 & 1,238,071 & 0.307 & 0.000 & 0.784\\
\bottomrule\end{tabular}

\caption{Shared model processing costs. B denotes bytes.}
\label{tab:agent-upstream}
\end{table}

We report pooled query correctness and balanced accuracy.
Abstentions count as incorrect for overall correctness.
Conditional error counts incorrect answers among answered queries.
Source disagreement compares a decoded answer with the decision obtained
directly from the original extracted categories.
We used complete-dialogue bootstrap resampling with seed 17102027.
The observation and corruption seed was 17102026.
Every query and outcome remains attached to its original dialogue.
The retained payloads, per-query outcomes, timings, and source probabilities
are included in the reproducibility package.

Table~\ref{tab:agent-storage} includes episode-specific archives and the
catalogue compiler's additional shared schema. Common extractor and
catalogue resources are listed separately in Table~\ref{tab:agent-protocol}.
The cohort-total column favors arithmetic mixed radix over the catalogue
statistic. A smaller private payload does not necessarily reduce total storage.
Let $D$ denote the number of stored dialogue states.
Let $p_{\mathrm{cat}}$ and $p_{\mathrm{mix}}$ denote their catalogue and
mixed-radix payload sizes, and let $S_{\mathrm{code}}$ denote the shared codebook size.
All sizes are in bytes. One unchanged codebook serves the complete cohort.
The serialized storage costs are
\begin{align}
 T_{\mathrm{cat}}(D)&=Dp_{\mathrm{cat}}+S_{\mathrm{code}},\\
 T_{\mathrm{mix}}(D)&=Dp_{\mathrm{mix}}.
\end{align}
Because $p_{\mathrm{mix}}>p_{\mathrm{cat}}$, the catalogue statistic uses less storage exactly when
\begin{equation}
 D>\frac{S_{\mathrm{code}}}{p_{\mathrm{mix}}-p_{\mathrm{cat}}}.
 \label{eq:amortization}
\end{equation}
The crossover rows in Table~\ref{tab:agent-storage} apply this expression
to the reported payload and codebook sizes.

\begin{table}[t]
\centering\small\begin{tabular}{lrrrr}\toprule
Method & State (B) & Archive (B) & Extra shared (B) & Cohort total (B)\\\midrule
Posterior summary & 312 & 0 & 0 & 311,688\\
Category vector & 7 & 0 & 0 & 6,993\\
Mixed radix & 3 & 0 & 0 & 2,997\\
Catalogue statistic & 2 & 0 & 8,878 & 10,876\\
Universal bit code & 4 & 4 & 0 & 7,992\\
Raw cache, 0 B & 0 & 4 & 0 & 3,996\\
Raw cache, 1 B & 1 & 4 & 0 & 4,995\\
Raw cache, 2 B & 2 & 4 & 0 & 5,994\\
Raw cache, 3 B & 3 & 4 & 0 & 6,993\\
Raw cache, 4 B & 4 & 0 & 0 & 3,996\\
\bottomrule\end{tabular}
\par\smallskip
\begin{tabular}{lr}\toprule
Catalogue/mixed-radix crossover & Stored states\\\midrule
Equal total storage & 8,878\\
Catalogue smaller from & 8,879\\\bottomrule\end{tabular}

\caption{Cohort storage and catalogue amortization, excluding common resources.
B denotes bytes.}
\label{tab:agent-storage}
\end{table}

Table~\ref{tab:agent-memory} gives the upstream process high-water resident
memory. The process includes dataset loading, models, text features,
and source-state construction. The measurement precedes allocation of
the result tensors used by the evaluation process.
We separately traced writer and reader allocations on sampled dialogues.
Those call measurements exclude pre-existing inputs, archives, public
codebooks, interpreter memory, and native library allocations.

\begin{table}[t]
\centering\small\begin{tabular}{lrr}\toprule
Observation source & Peak resident memory (B) & Allocation pairs\\\midrule
Native source & 223,133,696 & 200\\
History source & 260,161,536 & 200\\
\bottomrule\end{tabular}

\caption{Upstream resident memory and traced writer/reader call pairs.
B denotes bytes.}
\label{tab:agent-memory}
\end{table}

The computational environment used Python 3.12.12, NumPy 2.2.6,
SciPy 1.16.3, scikit-learn 1.7.2, and joblib 1.6.0.
Linear algebra used one thread. Timings followed a fixed method order
and describe the local reference implementation.

\section{Cached-output relation control}
We replayed the native Nous updater on cached text-classifier predictions
from MultiWOZ 2.4 \parencite{multiwoz24,nousrelease}.
We excluded one dialogue because its cached history was incomplete
for the required slots. The exclusion was independent of method success.
The updater reproduced every included slot decision.
We compared the native updater, the cached full-history classifier, and
the last local report. The full-history classifier receives a different
text interface and serves as a diagnostic comparison.

The final categorical slots admit a mixed-radix representation that
recovers every slot without retrieval. For the relation evaluation,
we represented every category except the last by a binary indicator.
We sampled pair questions using XOR, AND, OR, and equivalence, and
applied identical questions to each source method.
Table~\ref{tab:protocol} records the encoding sizes and sample counts.
Both encodings preserved every source answer.
The categorical representation used less retained storage and required
no archive, whereas the universal encoder covered the full binary cube.

Table~\ref{tab:dialogue} reports agreement with benchmark-derived labels.
We resampled complete dialogues to estimate the intervals.
The full-history classifier had higher accuracy in both query families.
The relation and slot columns measure different targets.
Categorical support explains the storage advantage: valid indicator
records occupy only a subset of the full cube.

\begin{table}[t]
\centering\small\begin{tabular}{lrrrr}\toprule
Source method & Slot accuracy & Relation accuracy & 95\% interval & Mismatches\\\midrule
Native Nous & 0.792 & 0.900 & [0.894, 0.906] & 0\\
Full-history classifier & 0.901 & 0.947 & [0.942, 0.952] & 0\\
Last local report & 0.772 & 0.891 & [0.884, 0.897] & 0\\
\bottomrule\end{tabular}

\caption{Source accuracy and encoding mismatches on dialogue-derived
records. Intervals are dialogue-bootstrap confidence intervals.}
\label{tab:dialogue}
\end{table}

\begin{table}[t]
\centering\small\begin{tabular}{ll}\toprule
Quantity & Value\\\midrule
Controlled record lengths (bits) & 10, 16, 64\\
Worlds per method-condition cell & 2,000\\
Method-condition cells & 96\\
Archive bit-corruption probabilities & 0.01, 0.10\\
Displayed controlled condition & $n=10$, $B=8$, $R\le1$, corruption $0.10$\\
Complete dialogues / included user turns & 999 / 7,360\\
Replayed slot decisions & 51,520\\
Slot radices & $(7,5,4,4,7,5,7)$\\
Possible categorical records & 137,200\\
Categorical retained state (bits / bytes) & 18 / 3\\
Binary indicators & 32\\
Universal retained state (bits / bytes) & 30 / 4\\
Universal binary-fact archive (bytes) & 32\\
Relation queries per dialogue & 64\\
Method-query executions & 191,808\\
Dialogue bootstrap replicates & 2,000\\
Shared extractor / text archive (bytes) & 29,791,103 / 13,790,771\\
Timing records per condition & 100\\
Allocation records per condition & 20\\
Raw write/query allocation pairs & 160\\
\bottomrule\end{tabular}

\caption{Evaluation protocol and physical resources.}
\label{tab:protocol}
\end{table}

\end{document}